\documentclass[a4paper]{article}

\usepackage{INTERSPEECH2022}
\usepackage[dvipsnames]{xcolor}

\title{UserLibri: A Dataset for ASR Personalization Using Only Text}
\name{Theresa Breiner,
Swaroop Ramaswamy,
Ehsan Variani,
Shefali Garg,
Rajiv Mathews,
Khe Chai Sim,
Kilol Gupta,
Mingqing Chen,
Lara McConnaughey}
\address{
  Google Inc.}
\email{\{tbreiner,swaroopram,variani,shefgarg,mathews,khechai,\\kilolgupta,mingqing,laramcc\}@google.com}

\begin{document}

\maketitle
\begin{abstract}
  Personalization of speech models on mobile devices (on-device personalization) is an active area of research, but more often than not, mobile devices have more text-only data than paired audio-text data. We explore training a personalized language model on text-only data, used during inference to improve speech recognition performance for that user. We experiment on a user-clustered LibriSpeech corpus, supplemented with personalized text-only data for each user from Project Gutenberg. We release this User-Specific LibriSpeech (UserLibri) dataset to aid future personalization research. LibriSpeech audio-transcript pairs are grouped into 55 users from the test-clean dataset and 52 users from test-other. We are able to lower the average word error rate per user across both sets in streaming and nonstreaming models, including an improvement of 2.5 for the harder set of test-other users when streaming.
\end{abstract}

\section{Introduction}

End-to-end (E2E) automatic speech recognition (ASR) systems can now run inference entirely on-device, thereby having important implications for latency, reliability and privacy \cite{he2019streamingE2E, sanaith2020streaming-surpassing-serverside}. The models themselves, which are typically trained server-side on large amounts of audio data, can perform quite well during inference on-device if the target domain is a reasonable match to the training data. However, they may perform poorly on rare words, or in applications where there is not a wealth of similar labeled training data available server-side. New research techniques are being developed to address this, such as by leveraging an external language model (LM) \cite{sainath2021efficient}.

On-device learning techniques such as Federated Learning (FL) and personalization (p13n\footnote{\label{p13n}In this paper, we use ``p13n'' to mean both ``personalization'' and ``personalized'', as in ``p13n LM''.}) can train models on the actual user data on-device without ever sending the raw user data to a central server. While a few active users of speech recognition services on mobile devices may have significant amounts of audio data on their device ($\approx40\%$ of the U.S. population uses a digital voice assistant at least once a month according to Statista \cite{voice-users-2021}), users tend to interact more with their mobile devices using virtual keyboards rather than speech recognition. Further, typical supervised training of ASR models requires ground-truth text labels for each utterance, which may not be available on mobile devices to use for p13n. However, by utilizing more of the on-device text data to learn user-preferred words or phrases, we can enable the on-device ASR models to possibly recognize them correctly the first time without access to any audio training examples containing that information.

In this work, we first generate a simulation dataset of \texttildelow100 users, based on the existing LibriSpeech dataset. We then explore the use of shallow fusion with personalized LMs to improve average word error rate (WER) per user on this data.

\section{Our Contributions}
\label{sec:contrib}

We present the UserLibri dataset\footnote{https://www.kaggle.com/datasets/google/userlibri}, a re-formatting of LibriSpeech containing paired audio-transcripts and extra text-only data, that can be useful in a variety of experiments. We hope that this dataset will help futher personalization research as to our knowledge there is no existing open-source dataset containing both user-specific audio and additional text-only personalized data. We report our initial experiments leveraging this dataset, which aim to improve WER of an ASR system via shallow fusion with personalized (p13n\footnotemark[1]) LMs. Our top findings are:

\begin{itemize}
\item p13n LM fusion can greatly improve WER, especially for certain users, and is better than non-p13n LM fusion,
\item streaming ASR models, which perform worse than non-streaming models, see greater absolute WER improvements with p13n LM fusion than non-streaming,
\item larger LMs perform better than smaller ones,
\item the number of LM training examples per user affects fusion performance, but even fine tuning a p13n LM on only 500 examples per user can beat the baseline,
\item p13n speech models can be combined with p13n LM fusion for even better results.
\end{itemize}

Background and related work motivating our research is in Section \ref{sec:background}. Dataset creation details are in Section \ref{sec:dataset}, followed by our models and experiments in Section \ref{sec:experiments}. We also include some specific wins and losses in Section \ref{sec:dive}. 

\section{Background}
\label{sec:background}

Machine learning models frequently have to handle the question of bias, where model performance on certain classes of data or certain users may be much worse than others, even despite excellent metrics on large test datasets \cite{mehrabi2021bias, martin2021habitual-be}.
Previous work has improved ASR performance for underrepresented classes of users via ASR p13n, by fine tuning a speaker-independent base model on a single speaker's audio examples \cite{green2021disordered, tomanek2021disordered}.

One specific type of data that can be difficult for natural language models is less common proper names or other tail words, which may not occur frequently in the training data and occur disproportionately depending on the user \cite{aleksic2015contact-names}.
Previous work has used shallow fusion \cite{gulcehre2015shallow} to leverage larger text corpora and combined it with biasing techniques that can target improvements for contexts frequently followed by proper nouns, such as ``call" and ``message" \cite{Zhao2019ShallowFusionEC}, or even without such context constraints \cite{le2021contextualized}.
Recent work has specifically explored personalizing end-to-end ASR models on-device by incorporating a biased LM on-the-fly, or by model fine-tuning on synthesized speech \cite{sim2019wikinames}. These approaches only require the target proper nouns in text form, possibly in a user-provided list. By contrast, we aim to improve ASR performance via personalization of an RNN LM requiring no active input from the user at all, and no real or synthesized user-specific audio data.

There are several ways of using LMs to improve ASR. One can interpolate the scores with the ASR model score, use them during the beam search, or use the LM as a rescoring model after the beam search. In this work, we choose to always interpolate the language model scores prior to the beam search, which is known as shallow fusion \cite{gulcehre2015shallow}.

Since one of the advantages of shallow fusion is the ability to leverage a much larger text dataset to improve an ASR model, most literature explores fusing with large LMs trained on the order of a million examples \cite{kannan2018externalLM}. However, in the on-device LM p13n setting, both training data and model size are much more limited.
Streaming models are preferred in the on-device setting due to their low latency at inference, despite their generally worse performance compared to nonstreaming models, which can analyze the entire input sequence before deciding the output.
Recent work showed that shallow fusion of an external RNN LM can be successful with streaming models, again training on hundreds of millions or even billions of examples \cite{cabrera2021streamingfusion}.

Related work in personalizing LMs has often focused on the dilemma of mismatched domains and sparseness of personalized text data, and incorporates adaptation or transfer learning techniques during p13n \cite{wen2013p13nlm}. While our simulation datasets all draw from matching text domains, the work in this paper introduces a new challenge of evaluating p13n LMs not on text-only tasks, where optimizing LM perplexity may suffice, but on ASR. To our knowledge, there is limited literature discussing this topic in the realm of p13n, although there are many works on optimizing or adapting LMs for speech tasks \cite{mikolov2011extensions, li2018conversational, deena2019broadcast}.

\section{User Specific LibriSpeech Dataset}
\label{sec:dataset}

LibriSpeech \cite{librispeech} is a widely-used dataset containing 970 hours of paired audio-transcript data, which are recordings of various speakers reading aloud from Project Gutenberg e-books \cite{gutenberg}. The Project Gutenberg (PG) raw book text data which served as the source for these recordings can also be downloaded from the LibriSpeech resources at https://www.openslr.org/12. The raw metadata files are also available which map audio examples to the corresponding PG source book texts.

The audio recordings are based on only a subset of text from each book, and the rest of the book text will contain similar vocabulary, character names, and writing style as the recordings. Therefore, we can create a dataset for a "user" containing the original paired audio-transcripts from a book as well as additional text-only data from the remainder of the book, which will match the user's domain. This multi-user dataset can be used for a variety of personalization applications beyond shallow fusion with personalized LMs, such as data augmentation studies using Text-to-Speech to create more audio training examples for a user \cite{rosenberg2019tts4asr}. While the LibriSpeech audio data is stored by speaker ID and chapter ID, we combine audio examples from chapters of the same book, read by the same speaker, into a single ``user" dataset. This results in more data per user, with an average of 52 audio examples per user (see more in Table \ref{tab:libri_data}).

\begin{table}[th]
  \caption{UserLibri dataset utterance \& p13n LM text metadata}
  \label{tab:libri_data}
  \centering
  \begin{tabular}{ l r r }
    \toprule
    \multicolumn{1}{c}{\textbf{Metadata}} &
    \multicolumn{1}{c}{\textbf{Test-Other}} &
    \multicolumn{1}{c}{\textbf{Test-Clean}} \\
    \midrule
    $\#$~Users                        & $52$ & $55$~~~             \\
    \midrule
    \textbf{Audio-Transcript Utts} & &\\
    Avg. $\#$ per User        & $56.5$  & $47.1$~~~               \\
    Median $\#$ per User         & $52$  & $46$~~~       \\
    $\#$ Users with $\geq$10    & $50$  & $52$~~~       \\
    Max $\#$ for $1$ User   & $144$  & $108$~~~              \\
    \midrule
    \textbf{LM Train Text Sentences} & & \\
    Total $\#$     & $444,520$  & $377,049$~~~      \\
    Avg. $\#$ per User & $8,548$  & $6,855$~~~    \\
    Median $\#$ per User  & $5,299$  & $3,750$~~~              \\
    $\#$ Users with $\geq$3k  & $43$  & $38$~~~              \\
    Max $\#$ for $1$ User    & $38,498$  & $54,306$~~~       \\
    \bottomrule
  \end{tabular}
  
\end{table}

To process the raw PG book text files into LM train sets of sentences, the text file encodings are standardized and boilerplate is removed. Then, the book text is broken into sentences, ignoring newlines mid-sentence that exist for readability in the raw data. For each sentence, we remove word-leading or trailing punctuation and sentence-leading or trailing whitespace, and uppercase the ascii characters. We then discard any sentences containing a sequence of $80\%$ of the sentence tokens where the same sequence can be found in the test audio examples which came from the same book.

In the UserLibri dataset, we only consider the users found in the LibriSpeech test-clean (considered easier for ASR) and test-other (considered harder/more noisy for ASR) datasets, as this allows us to use the LibriSpeech train and dev sets to train the ASR model that we use for our fusion experiments without training on any of the test speakers' data. We note that although LibriSpeech's train, dev, and test audio sets have no overlap in speakers, there are a few speakers in each set who read chapters from the same book, so the ASR model does see some text snippets from these books, read by different speakers. See Appendix \ref{app:process} for further details on the dataset generation process.

While we focus on the English LibriSpeech here, the Multilingual LibriSpeech could be similarly processed in the future.

\section{Experiments}
\label{sec:experiments}

We use the UserLibri dataset to explore whether shallow fusion with p13n LMs can improve per-user WER, as well as the effects of fusing with streaming ASR models vs. nonstreaming, and using different LM sizes trained on various amounts of data. We then combine p13n LM fusion with a p13n ASR model.

\subsection{Models}
\label{sec:models}

We train 86M parameter Conformer Hybrid Autoregressive Transducer (HAT) models \cite{variani2020hat} on the 960 hours of audio train data in LibriSpeech.
The audio data is preprocessed similarly to \cite{gulati2020conformer}, by extracting 80-channel filterbanks features computed from a 25ms window with a stride of 10ms. We use SpecAugment \cite{park2019specaug} with mask parameter $(F = 27)$, and ten time masks with maximum time-mask ratio ($p_S$ = 0.05), where the maximum-size of the time mask is set to $p_S$ times the length of the utterance.
Our models consist of 12 encoder layers with dimension 512, 4 attention heads, a convolution kernel size of 32, and a HAT decoder like in \cite{variani2020hat} with a single RNN layer of dimension 640. Each label is embedded with 128 dimensions, and inputs are tokenized with a 1k Word-Piece Model (WPM) trained on the LibriSpeech text-only data.
The models are trained with Adam \cite{kingma2015adam} and use GroupNorm \cite{wu2018groupnorm}.

We train two Conformer HAT models, one with streaming where the right context is 0 (no lookahead) which uses causal convolution and local self attention, and one nonstreaming with multi-headed attention.
To make sure that the encoder is streaming, we remove all the sub-architecture components which
were benefiting from right context. Namely, we remove the convolution
sub-sampling layer and also force the stacking layers to only stack
within the left context.

For both models, we always use time-synchronous beam search during decoding, resulting in slightly different baseline WERs than similar studies like \cite{gulati2020conformer}, which uses label-synchronous beam search and a right context of 1 frame.

We experiment with three RNN LM sizes with LSTM cells, all using the same 1024 WPM as the Conformer. The 3M parameter model uses 192 embeddings and 1 RNN layer of size 670. The 10M and 25M models use 384 embeddings and an RNN of size 1340, with 1 and 2 layers respectively.
We train general non-p13n LMs on the LibriSpeech LM text data (40M examples)\footnote{See Appendix \ref{app:genlm} for alternative general LM results.} using Adam with a learning rate of 1e-4, and batch size 4096. We then fine tune all weights for 1k steps on the user's p13n LM data, with batch size 32.

\subsection{Room for Improvement with Personalization}
\label{sec:improve}

We ran hyperparameter sweeps over the ASR models above to select the best top-K and beam width for the beam search, and the best smoothing temperature for the baseline models without shallow fusion (baseline-1 or BL1). Additionally, we swept over the HAT decoding hyperparameters from \cite{variani2020hat} - the internal LM weight, the external LM weight, and internal LM smoothing temperature - for the same models fused with a 10M general LM, trained on the entire LibriSpeech LM text corpus (BL2).

\begin{table}[th]
  \caption{Baseline WERs, for test-clean (CL) and test-other (OT), without p13n LM fusion. Average WERs per user (with 95$\%$ CIs) are higher than the WER on the full set of users.}
  \label{tab:baselines}
  \centering
  \begin{tabular}{ l l r r }
    \toprule
    \multicolumn{1}{c}{\textbf{Model}} &
    \multicolumn{1}{c}{\textbf{Set}} &
    \multicolumn{1}{c}{\textbf{Full Set}} &
    \multicolumn{1}{c}{\textbf{Average WER}}\\
    & & \multicolumn{1}{c}{\textbf{WER}} & \multicolumn{1}{c}{\textbf{per User}} \\
    \midrule
    \textbf{Streaming} & & & \\
    No Fusion (BL1) & CL & $5.8$ &  $6.0$ $[5.3, ~~~6.7]$ \\
    & OT & $10.4$ & $11.2$ $[9.6, ~13.0]$ \\
    Gen. 10M LM (BL2) & CL & $5.3$  & $\mathbf{5.4 ~[4.8, ~~6.0]}$ \\
    & OT & $9.0$ & $\mathbf{9.7 ~[8.3, 11.1]}$ \\
    \midrule
    \textbf{Nonstreaming} & & & \\
    No Fusion (BL1) & CL & $2.4$ &  $2.5$ $[2.1, ~~2.8]$ \\
    & OT & $5.8$ & $6.8$ $[5.6, ~~8.1]$ \\
    Gen. 10M LM (BL2) & CL & $2.1$  & $\mathbf{2.1~[1.8, ~2.5]}$ \\
    & OT & $5.0$ & $\mathbf{5.8~[4.8, ~6.9]}$ \\
    \bottomrule
  \end{tabular}
  
\end{table}

\begin{figure}[t]
  \centering
  \includegraphics[width=\linewidth]{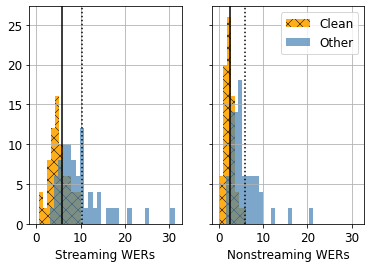}
  \caption{Histogram of per-user baseline WERs (no fusion). Baseline WER for the full test-clean is shown as a solid vertical line, and for test-other as a dotted vertical line. Many users have much higher baseline WERs, and p13n may help.}
  \label{fig:baseline_wers}
\end{figure}

With the best hyperparameters for the streaming and the nonstreaming models, both with and without shallow fusion using this non-p13n LM, we get the WERs reported in Table \ref{tab:baselines}. 95$\%$ confidence intervals (CIs) for averages throughout the paper are calculated using the bootstrap technique \cite{bisani2004bootstrap}, sampling the per-user WERs with replacement 10k times.

Our models achieve similar results on the full LibriSpeech test-other and test-clean sets compared to other studies such as \cite{gulati2020conformer} (note the differences as described in Section \ref{sec:models}). However, average per-user WERs are worse than WERs of the full set, especially in the test-other users, and even in the nonstreaming model despite its generally lower WERs. We hope to improve performance for the high-WER users (Figure \ref{fig:baseline_wers}).

Fusion with the general LM (BL2) improves performance, but still shows higher per-user WER averages than on the full sets. We will now describe how further per-user improvements can be made using p13n LMs.

\subsection{Results}
\label{sec:results}

\setlength{\tabcolsep}{3.5pt}

\begin{table}[th]
  \caption{Average WERs and 95$\%$ CIs from fusing with general (Gen) and p13n LMs of different sizes, for streaming (Str.) and nonstreaming (Nonstr.). BL1 is the baseline with no fusion.}
  \label{tab:fullp13n}
  \centering
  \begin{tabular}{ l r r r }
    \toprule
    \multicolumn{1}{c}{\textbf{Model}} &
    \multicolumn{1}{c}{\textbf{Test-Clean}} & 
    \multicolumn{1}{c}{\textbf{Test-Other}} &
    \multicolumn{1}{c}{\textbf{All Users}} \\
    \midrule
    \textbf{Str.} & & & \\
    BL1 & $6.0$ $[5.3, ~6.7]$ &  $11.2$ $[9.6, ~13.0]$ & $8.5$ $[7.5, ~9.6]$ \\
    3M Gen & $5.8$ $[5.1,~ 6.4]$ & $10.5$ $[9.0, ~12.0]$ & $8.1$ $[7.2, ~9.0]$ \\
    3M p13n & $5.6$ $[4.9, ~6.2]$ & $10.0$ $[8.4, ~11.6]$ & $7.7$ $[6.8, ~8.7]$ \\
    10M Gen & $5.4~[4.8, ~6.0]$ & $9.7$ $[8.3, ~11.1]$  & $7.5$ $[6.7, ~8.4]$ \\
    10M p13n & $5.4$ $[4.8, ~6.1]$ & $9.4~[8.0,~ 10.9]$ & $7.4~[6.5, ~8.3]$ \\
    25M Gen & $\mathbf{5.2~[4.7, 5.9]}$ & $9.1$ $[7.8, ~10.4]$ & $7.1$ $[6.3,~ 8.0]$ \\
    25M p13n & $\mathbf{5.2~[4.5, 5.9]}$ & $\mathbf{8.7~[7.4, 10.2]}$ & $\mathbf{6.9~[6.1, 7.8]}$ \\
    \midrule
    \textbf{Nonstr.} & & & \\
    BL1 & $2.5$ $[2.1, ~2.8]$  & $6.8$ $[5.6, ~8.1]$ &  $4.5$ $[3.8, ~5.4]$    \\
    3M Gen & $2.3$ $[2.1, ~2.7]$ & $6.4$ $[5.3, ~7.6]$ & $4.3$ $[3.6, ~5.1]$ \\
    3M p13n & $2.0$ $[1.7, ~2.4]$ & $5.8$ $[4.8, ~7.0]$ & $3.9$ $[3.2, ~4.5]$ \\
    10M Gen & $2.1$ $[1.8, ~2.5]$  &$5.8$ $[4.8, ~6.9]$  & $3.9$ $[3.3, ~4.6]$\\
    10M p13n & $\mathbf{1.9~[1.6, 2.3]}$ & $5.3$ $[4.3, ~6.4]$ & $3.6$ $[3.0, ~4.2]$ \\
    25M Gen & $2.0$ $[1.7, ~2.3]$ & $5.5$ $[4.6, ~6.6]$ & $3.7$ $[3.2, ~4.4]$ \\
    25M p13n & $\mathbf{1.9~[1.6, 2.3]}$ & $\mathbf{4.6~[3.8, 5.6]}$ & $\mathbf{3.2~[2.7, 3.8]}$ \\
    \bottomrule
  \end{tabular}
  
\end{table}

We sweep external LM weights of [0.15, 0.22, 0.36, 0.45, 0.55] for p13n LM shallow fusion with both the streaming and nonstreaming ASR models and notice that values above 0.22 quickly perform worse than the baseline, and for most users, 0.15 performs better than 0.22. We fuse using LM weight 0.15 for all users and leave the question of p13n acceptance criteria, or if/how to fuse per user, for future work. Results can be found in Table \ref{tab:fullp13n}, with p13n LMs outperforming the general LMs in almost every category. While the streaming model shows larger effects, even the nonstreaming model benefits from p13n LM fusion. Fusing with the 3M model already surpasses the baselines, with greater gains as model size increases. Per-User WER histograms are included in Appendix \ref{app:plots}.

\begin{figure}[t]
  \centering
  \includegraphics[width=\linewidth]{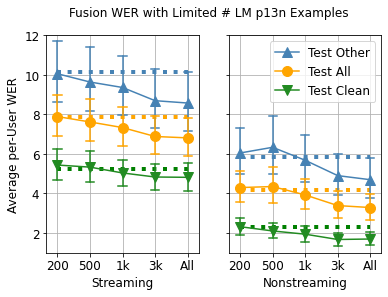}
  \caption{Average WER per user (and 95$\%$ CI) for p13n 10M LM fusion on limited user LM train sets, for Test-Clean, Test-Other and all users combined, for the 81 users with at least 3k LM train examples. Horizontal dotted lines show the baseline WERs (no fusion) for each group.}
  \label{fig:limited_train}
\end{figure}

We also experiment with fine tuning the p13n LMs on varying amounts of user LM examples: train sets of 200, 500, 1k, and 3k examples per user (see Appendix \ref{app:limiting}), which we compared to fine tuning on the full amount of LM data we have available for that user, only for the 81 users that have at least 3k examples total.
In Figure \ref{fig:limited_train}, we see that training with 3k examples performs almost as well as using all available data, and we can beat the baselines as long as we personalize on at least 500 examples for streaming models and 1k for nonstreaming. 

\subsection{Interaction with Speech Personalization}
\label{sec:speechp13n}

As mentioned in Section \ref{sec:background}, speech p13n by fine tuning full or partial models on-device can greatly improve the recall of proper nouns such as named entities \cite{sim2019wikinames}. We briefly try combining speech p13n and p13n LM fusion\footnote{We don't sweep any hyperparameters for these speech experiments.}.

We fine tune the joint layer of the streaming ASR model for each user via 5-fold cross validation, for the 102 users with at least 10 audio-transcript pairs, and average the per-fold WERs. Speech p13n alone improves upon the baseline, with non-p13n speech + p13n LM fusion performing a little better. Combining the two approaches gets the best results (Table \ref{tab:speech_p13n}).

\setlength{\tabcolsep}{3pt}

\begin{table}[th]
  \caption{Average WERs from personalizing the streaming ASR model on each user's test utterances via 5-fold cross validation, with and without p13n LM fusion. No p13n means no speech p13n or p13n LM fusion (BL1). Only the 102 users with at least 10 audio-transcript pairs are included. Reports $95\%$ CIs.}
  \label{tab:speech_p13n}
  \centering
  \begin{tabular}{ l r r r }
    \toprule
    \multicolumn{1}{c}{\textbf{Model}} &
    \multicolumn{1}{c}{\textbf{Test-Clean}} &
    \multicolumn{1}{c}{\textbf{Test-Other}} &
    \multicolumn{1}{c}{\textbf{All Users}} \\
    \midrule
    No p13n & $6.0$ ~$[5.3, ~6.6]$ &  $11.3$~ $[9.6, ~13.1]$ & $8.6$~ $[7.6, ~9.7]$ \\
    +10M LM & $5.6$ ~$[5.0, ~6.3]$ & $9.3$ ~$[7.9, ~10.8]$ & $7.4$~ $[6.6, ~8.3]$ \\
    p13n ASR & $5.7$ ~$[5.1,~ 6.3]$ & $10.4$ ~$[9.0, ~12.1]$ & $8.0$~ $[7.1, ~9.0]$ \\
    +10M LM & $\mathbf{5.4~[4.8, 6.0]}$ & $\mathbf{8.7~[7.5, 10.0]}$ & $\mathbf{7.0 ~[6.2, 7.8]}$ \\
    \bottomrule
  \end{tabular}
  
\end{table}

\subsection{Wins and Losses}
\label{sec:dive}

If we examine specific cases where the p13n LM is able to correctly predict examples that the baseline or general LM predicted incorrectly (Table \ref{tab:wins}), we see that proper nouns seen more frequently in the p13n LM training data can be a source of the win, but there are also some losses due to over-predicting words seen in the p13n LM data. Combining speech and LM p13n may allow wins on tail words while avoiding losses on small common words (Table \ref{tab:speech_wins}). See Appendix \ref{app:more_wins} for more.

\begin{table}[th]
  \caption{Example predictions. p13n LM fusion can win (W) on names, but may lose (L) if a test word rarely appears in the rest of the p13n LM data, but a similar word appears many times.}
  \label{tab:wins}
  \centering
  \begin{tabular}{l l l l l}
    \toprule
    \multicolumn{1}{c}{\textbf{W/L}} &
    \multicolumn{1}{c}{\textbf{Baseline}} &
    \multicolumn{1}{c}{\textbf{General}} &
    \multicolumn{1}{c}{\textbf{P13n}} &
    \multicolumn{1}{c}{\textbf{Count in}} \\
    & & \multicolumn{1}{c}{\textbf{LM}} & \multicolumn{1}{c}{\textbf{LM}} &
    \multicolumn{1}{c}{\textbf{p13n data}} \\
    \midrule
    W &king & king & king & sharkan: 0\\
    & \textcolor{BrickRed}{sharkan} &  \textcolor{BrickRed}{sharkan} & \textcolor{ForestGreen}{\textbf{sharrkan}} & sharrkan: 336 \\
    \midrule[1pt]
    W & \textcolor{BrickRed}{mardock}  & \textcolor{BrickRed}{murdock}  & \textcolor{ForestGreen}{\textbf{murdoch}}  & m[a/u]rdock: 0\\
    & blinked &  blinked & blinked & murdoch: 92\\
    \midrule[1pt]
    W & thanks  & thanks  & thanks & is he: 72\\
    & \textcolor{BrickRed}{is he} & \textcolor{BrickRed}{is he} & \textcolor{ForestGreen}{\textbf{izzy}} & izzy: 155\\
    \midrule[1pt]
    L & on the & on the & on the & navel: 0 \\
    & \textcolor{ForestGreen}{navel} & \textcolor{ForestGreen}{navel} & \textcolor{BrickRed}{\textbf{naval}} & naval: 2 \\
    \midrule[1pt]
    L & \textcolor{ForestGreen}{tied} to & \textcolor{ForestGreen}{tied} to & \textcolor{BrickRed}{\textbf{tide}} to & tied: 0\\
    & a woman & a woman & a woman & tide: 4 \\
    \bottomrule
  \end{tabular}
  
\end{table}

\begin{table}[th]
  \caption{Predictions on one utterance for speech p13n combined with LM p13n, which may allow similar tail word wins from Table \ref{tab:wins} while avoiding losses on smaller common words.}
  \label{tab:speech_wins}
  \centering
  \begin{tabular}{l l}
    \toprule
    \multicolumn{1}{c}{\textbf{Experiment}} &
    \multicolumn{1}{c}{\textbf{Prediction}}  \\
    \midrule
    Baseline & \textcolor{BrickRed}{ten years} farewell vintage is \textcolor{BrickRed}{none} \\
    General LM & \textcolor{BrickRed}{ten years} farewell vintage is \textcolor{ForestGreen}{done} \\
    p13n LM only & \textcolor{ForestGreen}{panniers} farewell vintage is \textcolor{BrickRed}{none} \\
    Speech p13n only & \textcolor{ForestGreen}{panniers} farewell vintage is \textcolor{BrickRed}{none} \\
    Speech \& LM p13n & \textbf{\textcolor{ForestGreen}{panniers}} farewell vintage is \textbf{\textcolor{ForestGreen}{done}} \\
    \bottomrule
  \end{tabular}
  
\end{table}

\section{Conclusions}
\label{sec:conclusion}

We share our new personalization simulation dataset, UserLibri, based on LibriSpeech and useful for a variety of personalization research. We demonstrate a use case where we are able to improve average per-user WER in both streaming and nonstreaming models, bringing down streaming WER on the hardest group of users by 1.2 by performing shallow fusion with personalized LMs of only 3M parameters, or an improvement of 2.5 with 25M LMs. We show that the technique can still work with limited p13n LM examples and can be combined with speech p13n for even better results.

\clearpage

\bibliographystyle{IEEEtran}

\bibliography{mybib}

\newpage
\appendix

\onecolumn
\begin{center}
    {\Large{UserLibri Appendix}} \\
    {\Large{Interspeech 2022 Submission, Breiner et. al.}}
\end{center}

\section{Processing the Raw Book Data}
\label{app:process}

While Project Gutenberg (PG) is a highly useful and well organized data resource, there are some inconsistencies that come along with such a large crowd-sourced dataset that we had to handle during the UserLibri dataset generation. We were only processing the 99 books that were included as transcript sources for the LibriSpeech test-clean and test-other data as mentioned in Section \ref{sec:dataset}, so we only know of the fixes we had to make for that small subset of the PG corpus.

The PG book texts can be downloaded from the LibriSpeech resources site at https://www.openslr.org/12. Most of the book text files are named with their book ID number as referenced in the BOOKS.txt metadata file that is provided, but some book texts are not named that way, although they do appear in a directory correctly named with the book ID. To ease processing, we first locally renamed all of the book texts to match the directory name.

We then had to standardize the file encodings. Most of the PG book text files are encoded in ascii format, with some encoded in UTF-8 which appear in a separate directory. However, some of the files are encoded in Windows-1252 encoding and were not readable by our pipelines until we first converted to UTF-8. Namely, we noticed this problem with the following book IDs: 19019, 3436, 3440, 3441.

Within each book text, there were also various boilerplate conventions giving the metadata about PG, the book, the licensing information or other non-book-content information, which we did not want to include in our LM train examples. We found the majority to use variations on simple lines such as ``***START OF THIS PROJECT GUTENBERG EBOOK'' and ``*** END OF THIS PROJECT GUTENBERG EBOOK'' to demarcate the beginning and end of the actual book content, with small differences such as ``THE'' instead of ``THIS'' or varying numbers of asterisks. A minority of the books used similar boilerplate although in lowercase, or with a different demarcation for the end of the book, while a couple had no boilerplate at all. We were able to leverage regular expressions to remove the boilerplate for our 99 target books at scale.

We note in Section \ref{sec:dataset} that while the speakers are properly unique across LibriSpeech train, dev and test set splits and only appear in one of these splits, the book text sources are not kept separate in this way. One book often serves as the source of transcripts for more than one speaker, and the different speakers may appear in different data splits. Specifically, 87 books are read by different speakers to create audio data in both the train and test sets, 66 books in both dev and test, and 98 books in both train and dev. This should not have a large effect on our experiments, but it does mean that a character name or other tail words seen in a user's p13n LM data (pulled from LibriSpeech test) may still occur in the LibriSpeech audio and LM examples used to train the ASR model and the General LM. However, these long tail words would appear rarely in the ASR and General LM train data, and disproportionately frequently in the p13n LM data and the test utterances, which is why we are still able to see gains with p13n LM fusion.

\section{Additional General LM Fusion Experiments}
\label{app:genlm}
In our main paper, we only report two baselines for fusion without p13n LMs, BL1 with no fusion and BL2 where we fuse with a 10M general LM trained on all LibriSpeech LM data (Table \ref{tab:baselines}). We explored a third baseline as well, where we trained a general 10M LM on only the LM train data from our 107 users that were generated in Section \ref{sec:dataset}, combined into a single set (650k examples). While there are a few books that serve as the p13n data for more than one user and thus appear multiple times in the full UserLibri set, these examples are only included once when training this smaller LM. Training an LM on this set parallels the type of LM we might be able to train if we have no central in-domain data and use a technique such as FL to train only on the text found on-device. We call this BL3 and compare results in Table \ref{tab:app_baselines}.

\begin{table}[th]
  \caption{Baseline WERs, for test-clean (CL) and test-other (OT), without p13n LM fusion. Average WERs per user, with 95$\%$ confidence intervals (CIs), are higher in almost every case than the WER on the full set.}
  \label{tab:app_baselines}
  \centering
  \begin{tabular}{ l l r r }
    \toprule
    \multicolumn{1}{c}{\textbf{Model}} &
    \multicolumn{1}{c}{\textbf{Set}} &
    \multicolumn{1}{c}{\textbf{Full Set}} &
    \multicolumn{1}{c}{\textbf{Average WER}}\\
    & & \multicolumn{1}{c}{\textbf{WER}} & \multicolumn{1}{c}{\textbf{per User}} \\
    \midrule
    \textbf{Streaming} & & & \\
    No Fusion (BL1) & CL & $5.8$ &  $6.0$ $[5.3, ~~~6.7]$ \\
    & OT & $10.4$ & $11.2$ $[9.6, ~13.0]$ \\
    General 10M LM (BL2) & CL & $5.3$  & $\mathbf{5.4 ~[4.8, ~~6.0]}$ \\
    & OT & $9.0$ & $\mathbf{9.7 ~[8.3, 11.1]}$ \\
    User-Only 10M LM (BL3) & CL & $5.5$ & $5.7$ $[5.0, ~~~6.3]$ \\
    & OT & $9.4$ & $10.2$ $[8.7, ~11.9]$ \\
    \midrule
    \textbf{Nonstreaming} & & & \\
    No Fusion (BL1) & CL & $2.4$ &  $2.5$ $[2.1, ~~2.8]$ \\
    & OT & $5.8$ & $6.8$ $[5.6, ~~8.1]$ \\
    General 10M LM (BL2) & CL & $2.1$  & $\mathbf{2.1~[1.8, ~2.5]}$ \\
    & OT & $5.0$ & $\mathbf{5.8~[4.8, ~6.9]}$ \\
    User-Only 10M LM (BL3) & CL & $2.4$ & $2.4$ $[2.0, ~~2.8]$ \\
    & OT & $4.9$ & $\mathbf{5.8 ~[4.8,~ 7.0]}$ \\
    \bottomrule
  \end{tabular}
  
\end{table}

Fusion with the general LM (BL2) gives best improvements overall, but even fusing with the LM which was only trained on the far smaller set of user p13n text data (BL3) improves upon the baselines, especially when streaming. However, all baselines still show higher per-user WER averages than on the full set of combined users. For the remainder of our experiments, we chose to focus on personalizing the general LM which trained on the full dataset (BL2) as its baseline performance is best, and leave personalizing a model trained only with the union of user-specific text examples to future work.

\section{Per-User WER Histograms}
\label{app:plots}

\begin{figure}[t]
  \centering
  \includegraphics[scale=0.6]{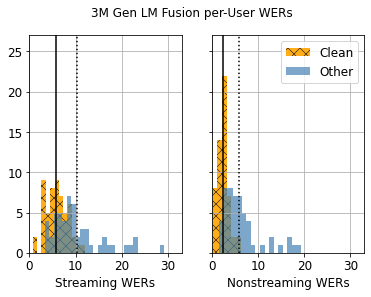}
  \includegraphics[scale=0.6]{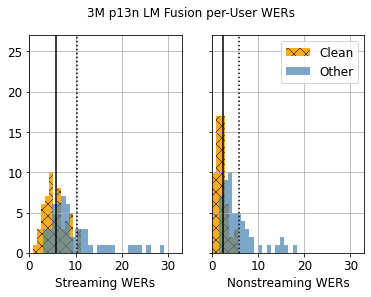}
  \caption{Histograms of per-user WERs with 3M General and p13n LM fusion.}
  \label{fig:3m_fusion_hist}
\end{figure}

\begin{figure}[t]
  \centering
  \includegraphics[scale=0.6]{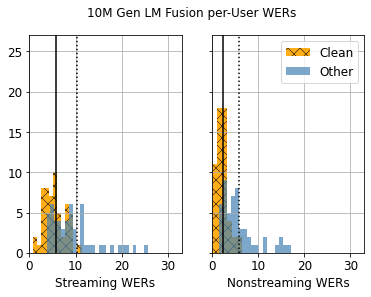}
  \includegraphics[scale=0.6]{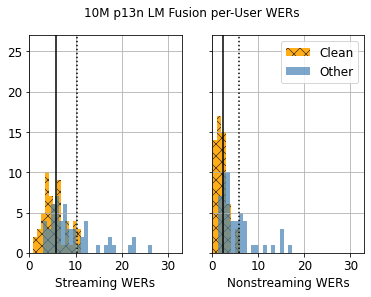}
  \caption{Histograms of per-user WERs with 10M General and p13n LM fusion.}
  \label{fig:10m_fusion_hist}
\end{figure}

\begin{figure}[t]
  \centering
  \includegraphics[scale=0.6]{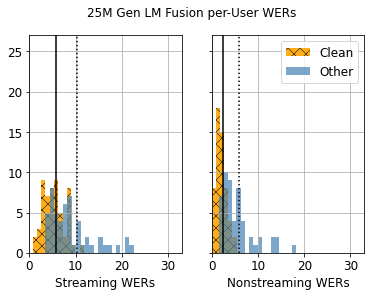}
  \includegraphics[scale=0.6]{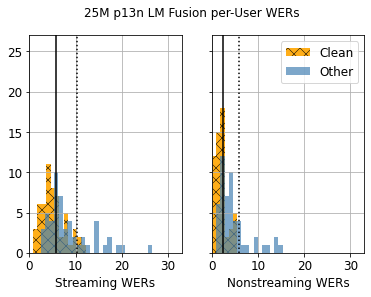}
  \caption{Histograms of per-user WERs with 25M General and p13n LM fusion.}
  \label{fig:25m_fusion_hist}
\end{figure}

To compare to Figure \ref{fig:baseline_wers} showing the baseline per-user WER distributions, some additional histograms showing the per-user WERs for various experiments are shown in Figures \ref{fig:3m_fusion_hist}, \ref{fig:10m_fusion_hist}, and \ref{fig:25m_fusion_hist}. The WER on the full sets of combined users are shown as vertical lines, solid for WER on test-clean and dotted for WER on test-other. We see the mass of each distribution, and particularly the high WER outliers on the right end of the x-axis, shift left from General LM fusion to p13n LM fusion, but also as we increase the model size from 3M to 10M to 25M parameters.

\section{Limited Training Examples Experiments}
\label{app:limiting}

In order to conduct the limited training examples experiments, we created new, separate LM datasets containing only 200, 500, 1k or 3k examples for each user. Each of these train sets is a strict subset of the next size set, so for example the 500 example set is created by taking the same 200 example set and adding 300 more randomly selected examples from the full p13n LM training data set for that user. This ensures that the experiments are more directly comparable and depend less on having included lucky examples in certain of the sets but not others.

\section{Additional Wins and Losses}
\label{app:more_wins}

In Table \ref{tab:more_wins} we show a larger set of examples building on Table \ref{tab:wins}, where p13n LMs are able to improve upon the baseline and General LM experiments, for example with long tail proper names, but may overfit to smaller common words in the p13n LM dataset and also cause some losses. In some examples, such as ``horse sense'', there are roughly equal occurrences in the p13n LM data of both the ground truth word and the predicted word, which may be why the p13n LM did not help there. However, for proper names the p13n still often is able to beat the baseline and general LM.

In Table \ref{tab:more_speech_wins} we show a few more examples not included in the main paper to compare p13n LM fusion with and without speech p13n. The wins that the speech p13n and combined speech p13n \& LM p13n see are mostly not on proper nouns, but often allow recovery from predictions that were correct with General LM fusion, incorrect with p13n LM fusion, but are then correct with the combination speech p13n and p13n LM fusion. Combining these approaches may allow the p13n LM fusion to show wins on proper nouns elsewhere in the user test data but not over-predict incorrect common words from the p13n LM data.

\begin{table}[th]
  \caption{Predictions per approach. p13n LM fusion is able to get wins (W) on proper names, but may lose (L) if a word only appears once in a test example and a similar word appears many times in the p13n LM data.}
  \label{tab:more_wins}
  \centering
  \begin{tabular}{l l l l l}
    \toprule
    \multicolumn{1}{c}{\textbf{W/L}} &
    \multicolumn{1}{c}{\textbf{Baseline}} &
    \multicolumn{1}{c}{\textbf{General}} &
    \multicolumn{1}{c}{\textbf{P13n}} &
    \multicolumn{1}{c}{\textbf{Count in}} \\
    & & \multicolumn{1}{c}{\textbf{LM}} & \multicolumn{1}{c}{\textbf{LM}} &
    \multicolumn{1}{c}{\textbf{p13n data}} \\
    \midrule
    W &king & king & king & sharkan: 0\\
    & \textcolor{BrickRed}{sharkan} &  \textcolor{BrickRed}{sharkan} & \textcolor{ForestGreen}{\textbf{sharrkan}} & sharrkan: 336 \\
    \midrule[1pt]
    W & lord of & lord of & lord of & bagdad: 0\\
    & \textcolor{BrickRed}{bagdad} & \textcolor{BrickRed}{bagdad} & \textcolor{ForestGreen}{\textbf{baghdad}} & baghdad: 71\\
    \midrule[1pt]
    W & mister & mister & mister & bell: 3\\
    & \textcolor{BrickRed}{bell} & \textcolor{BrickRed}{bell} & \textcolor{ForestGreen}{\textbf{beale}} & beale: 340\\
    \midrule[1pt]
    W & \textcolor{BrickRed}{mardock}  & \textcolor{BrickRed}{murdock}  & \textcolor{ForestGreen}{\textbf{murdoch}}  & m[a/u]rdock: 0\\
    & blinked &  blinked & blinked & murdoch: 92\\
    \midrule[1pt]
    W & thanks  & thanks  & thanks & is he: 72\\
    & \textcolor{BrickRed}{is he} & \textcolor{BrickRed}{is he} & \textcolor{ForestGreen}{\textbf{izzy}} & izzy: 155\\
    \midrule[1pt]
    W & a \textcolor{BrickRed}{peck} of  & a \textcolor{BrickRed}{peck} of  & a \textbf{\textcolor{ForestGreen}{bag}} & peck: 0\\
    & \textcolor{BrickRed}{fields} & \textcolor{BrickRed}{fields} & \textcolor{ForestGreen}{\textbf{eels}} & bag: 9\\
    \midrule[1pt]
    L & on the & on the & on the & navel: 0 \\
    & \textcolor{ForestGreen}{navel} & \textcolor{ForestGreen}{navel} & \textcolor{BrickRed}{\textbf{naval}} & naval: 2 \\
    \midrule[1pt]
    L & \textcolor{ForestGreen}{tied} to & \textcolor{ForestGreen}{tied} to & \textcolor{BrickRed}{\textbf{tide}} to & tied: 0\\
    & a woman & a woman & a woman & tide: 4 \\
    \midrule[1pt]
    L & nearer  & nearer  & nearer  & virtuous: 15\\
    & \textcolor{ForestGreen}{virtuous} &  \textcolor{ForestGreen}{virtuous} &  \textcolor{BrickRed}{\textbf{virtues}} & virtues: 24 \\
    \midrule[1pt]
    L & \textcolor{BrickRed}{poor}  & \textcolor{ForestGreen}{horse}  & \textbf{\textcolor{BrickRed}{poor}}  & poor: 7\\
    & \textcolor{ForestGreen}{sense} &  \textcolor{ForestGreen}{sense} &  \textcolor{BrickRed}{\textbf{cents}} & horse: 9 \\
    \bottomrule
  \end{tabular}
  
\end{table}

\begin{table}[th]
  \caption{Additional predictions using speech p13n with and without LM p13n, which may allow similar tail word wins from Table \ref{tab:more_wins} while avoiding losses on smaller common words.}
  \label{tab:more_speech_wins}
  \centering
  \begin{tabular}{l l}
    \toprule
    \multicolumn{1}{c}{\textbf{Experiment}} &
    \multicolumn{1}{c}{\textbf{Prediction}}  \\
    \midrule
    Baseline & \textcolor{BrickRed}{ten years} farewell vintage is \textcolor{BrickRed}{none} \\
    General LM & \textcolor{BrickRed}{ten years} farewell vintage is \textcolor{ForestGreen}{done} \\
    p13n LM only & \textcolor{ForestGreen}{panniers} farewell vintage is \textcolor{BrickRed}{none} \\
    Speech p13n only & \textcolor{ForestGreen}{panniers} farewell vintage is \textcolor{BrickRed}{none} \\
    Speech \& LM p13n & \textbf{\textcolor{ForestGreen}{panniers}} farewell vintage is \textbf{\textcolor{ForestGreen}{done}} \\
    \midrule
    Baseline & but this \textcolor{BrickRed}{our banishment} utterly \\
    General LM & but this \textcolor{BrickRed}{our banishment} utterly \\
    p13n LM only & but this \textcolor{BrickRed}{our vanity} utterly \\
    Speech p13n only & but this \textbf{\textcolor{ForestGreen}{shall banish it}} utterly \\
    Speech \& LM p13n & but this \textbf{\textcolor{ForestGreen}{shall banish it}} utterly \\
    \midrule
    Baseline & wars that have been \textcolor{BrickRed}{raged} \\
    General LM & wars that have been \textbf{\textcolor{ForestGreen}{waged}} \\
    p13n LM only & wars that have been \textcolor{BrickRed}{raged} \\
    Speech p13n only & wars that have been \textbf{\textcolor{ForestGreen}{waged}} \\
    Speech \& LM p13n & wars that have been \textbf{\textcolor{ForestGreen}{waged}} \\
    \midrule
    Baseline & \textcolor{BrickRed}{he does} never made man moral \\
    General LM & \textbf{\textcolor{ForestGreen}{it has}} never made man moral \\
    p13n LM only & \textcolor{ForestGreen}{it} \textcolor{BrickRed}{is} never made man moral \\
    Speech p13n only & \textbf{\textcolor{ForestGreen}{it has}} never made man moral \\
    Speech \& LM p13n & \textbf{\textcolor{ForestGreen}{it has}} never made man moral \\
    \midrule
    Baseline & \textcolor{BrickRed}{her legion} has been \textcolor{BrickRed}{cried} \\
    General LM & \textbf{\textcolor{ForestGreen}{religion}} has been \textbf{\textcolor{ForestGreen}{tried}}  \\
    p13n LM only & \textcolor{ForestGreen}{religion} has been \textcolor{BrickRed}{cried} \\
    Speech p13n only & \textcolor{BrickRed}{her legion} has been \textcolor{ForestGreen}{tried} \\
    Speech \& LM p13n & \textbf{\textcolor{ForestGreen}{religion}} has been \textbf{\textcolor{ForestGreen}{tried}} \\
    \bottomrule
  \end{tabular}
  
\end{table}

\end{document}